\RequirePackage{fix-cm}
\documentclass[natbib,smallextended]{svjour3}       
\smartqed  
\usepackage{graphicx}
\usepackage{upgreek}
 \journalname{Space Science Reviews}
\begin{document}

\title{Outstanding Challenges of Exoplanet Atmospheric Retrievals
}


\author{Joanna K. Barstow        \and
        Kevin Heng 
}


\institute{J.~K. Barstow \at School of Physical Sciences, The Open University, Walton Hall, Kents Hill, Milton Keynes, MK7 6AA, UK\at
              Department of Physics and Astronomy, University College London, Gower Street, London, WC1E 6BT, UK \\
              \email{jo.barstow@open.ac.uk}           
           \and
           K. Heng\at
          Center for Space and Habitability, University of Bern, Gesellschaftsstrasse 6, CH-3012 Bern, Switzerland. 
}

\date{Received: date / Accepted: date}

\maketitle

\begin{abstract}
Spectral retrieval has long been a powerful tool for interpreting planetary remote sensing observations. Flexible, parameterised, agnostic models are coupled with inversion algorithms in order to infer atmospheric properties directly from observations, with minimal reliance on physical assumptions. This approach, originally developed for application to Earth satellite data and subsequently observations of other Solar System planets, has been recently and successfully applied to transit, eclipse and phase curve spectra of transiting exoplanets. In this review, we present the current state-of-the-art in terms of our ability to accurately retrieve information about atmospheric chemistry, temperature, clouds and spatial variability; we discuss the limitations of this, both in the available data and modelling strategies used; and we recommend approaches for future improvement. 

\end{abstract}

\section{Introduction}
\label{intro}
Remote sensing of exoplanet atmospheres is a rapidly expanding field, having progressed from the first detections of a molecular species in an atmosphere around another star \citep{barman07,tinetti07} to beginning to characterize complex temperature structures, clouds, and spatial heterogeneity in just over ten years. Retrieval methods -- iteratively comparing synthetic to observed spectra in order to infer the most likely atmospheric state -- were historically applied to Solar System atmospheres, and with some adaptation are now being used to analyze exoplanet spectra. Typically, retrieval codes couple a simple, parameterised, 1D radiative transfer model to a retrieval algorithm. The model parameters form the atmospheric state vector, and the output from the retrieval algorithm is a posterior probability distribution for each element in the state vector, including correlations between the model parameters.

This review paper, rather than simply presenting an overview of the current state of the art, instead discusses what we see as the major challenges facing exoplanet retrievals over the next few years, and thus the directions in which we expect development to be most rapid. In general, all of these challenges can be summarized as resolving the tension between model realism (with risks of overfitting or allowing informative priors to drive solutions) and model simplicity (with the risk that the model may be inadequate to accurately reproduce the data, or may reproduce them for the wrong reasons, and may be very far from the truth). For each challenge, we present the current status, and then provide our recommendations for future routes of exploration and improvement. The key areas which we have identified are listed below:

\begin{enumerate}
    \item Inferring chemistry from measured molecular abundances
    \item Representation of temperature structure
    \item Representation of clouds and aerosols
    \item Including 3D effects in 1D models
\end{enumerate}

These areas will be dealt with in turn from Section~\ref{chemistry}. 

\subsection{Retrieval algorithms}
\label{algorithms}
\begin{figure}
    \centering
    \includegraphics[width=0.9\textwidth]{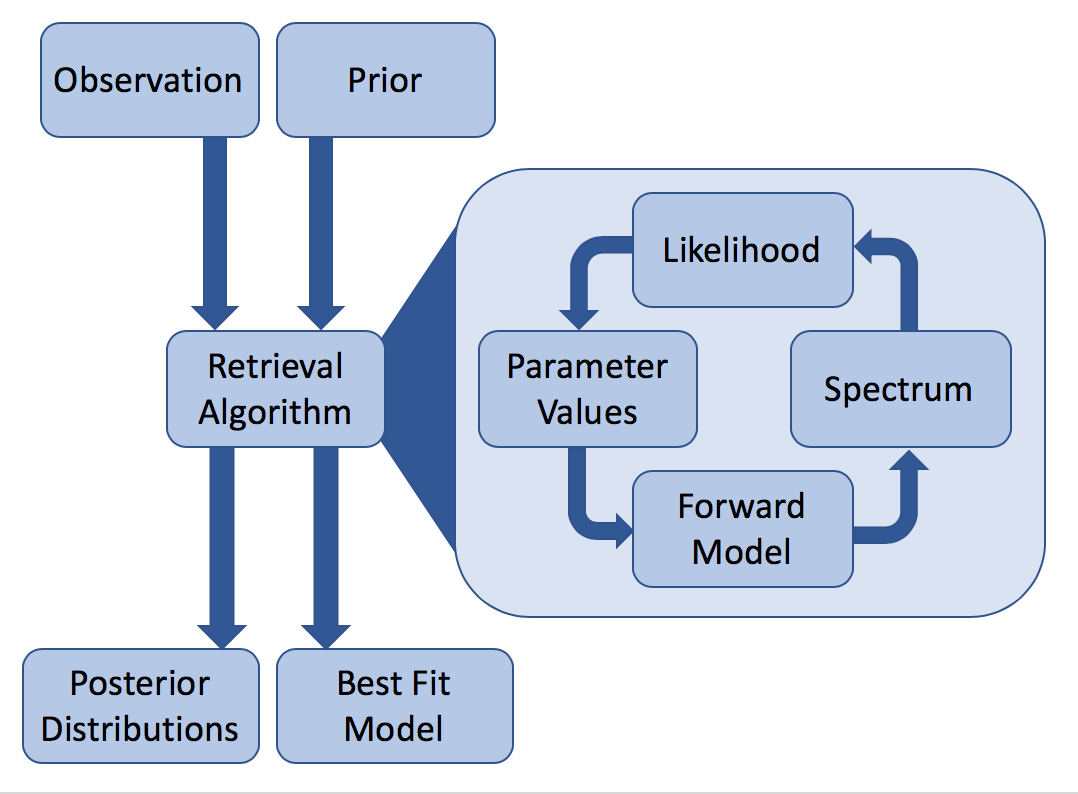}
    \caption{Schematic showing the basic structure of a retrieval algorithm.}
    \label{retrieval_schematic}
\end{figure}

A range of algorithms and retrieval codes have been applied to exoplanet retrievals, with each approach having different benefits. Figure~\ref{retrieval_schematic} shows the basic structure of a retrieval code. The earliest exoplanet retrievals used either a simple grid search (e.g. \citealt{madhu09}) or Optimal Estimation \citep{rodg00,irwin08}. Grid searches are simple to set up, but can be inefficient (since they may involve a detailed exploration of parameter space far from the solution) and results will be highly restricted by the parameter values included within the grid. Optimal Estimation is a matrix inversion method that assumes Gaussian statistics. A Levenberg-Marquardt scheme is used to iteratively solve the inverse problem and works to minimize a cost function, which assesses both the difference between the model output and the measured spectrum and also the distance of the atmospheric state vector from a Gaussian prior state vector.
Due to its imposition of Gaussianity, whilst Optimal Estimation is fast and efficient it is unable to a) effectively explore multimodal parameter spaces and b) explore a very broad parameter space, as the parameter ranges are restricted by the necessity of including a Gaussian prior constraint. 

Markov-chain Monte Carlo (MCMC; see e.g. \citealt{line13a}) and nested sampling algorithms have more recently become the preferred tools within the community. These Bayesian approaches both allow a more comprehensive exploration of the parameter space, as they do not restrict priors or posteriors to obeying Gaussian statistics. Of these approaches, the \verb{MultiNest{ \citep{feroz08,feroz09,feroz13} implementation of the nested sampling method \citep{skilling06} has proved especially popular, as this provides a relatively efficient exploration of potentially multi-modal posteriors that effectively captures multiple modes and complex degeneracies. 

With standard retrieval methods, there is always tension between achieving physical and chemical realism versus completing the calculation within a reasonable period of time.  In practice, this means that the forward models converting chemical abundances and opacities into transit radii and fluxes need to be simplified to enable rapid computation.  Recently, machine learning approaches have been adapted to performing atmospheric retrieval, which allows the burden of computing synthetic spectra to be shifted offline.  A grid of models may be computed beforehand and used as a training set for the supervised machine learning method of choice.  This approach has been demonstrated using regression trees and random forests \citep{marquez18}.  It allows model grids from different research groups to be used for atmospheric retrieval, even if the computer codes used to generate these models are proprietary and non-public.  Nevertheless, paying attention to model assumptions remains a key part of the process.  The unsupervised machine learning method of deep convolutional generative adversarial networks has also been implemented for atmospheric retrieval \citep{zingales18}.

\subsection{Parameterised 1D models}
\label{1Dmodels}

A general requirement for the majority of retrieval schemes is for the forward model computation to be fast. This is especially necessary for Monte Carlo and nested sampling methods, as these typically require millions of individual forward models to be computed to adequately explore the parameter space. Therefore, forward models must be relatively simple; instead of containing detailed physics, models are usually parameterised, and are generally also one-dimensional. Parameterisation must be approached with care, as simple models uncoupled from physical assumptions may be prone to converging on unrealistic solutions (e.g. atmospheres with implausible chemistry, or temperature-pressure profiles that would be unstable). However, this potential disadvantage can also be a strength in situations where our understanding of the underlying physics and chemistry is still relatively immature, as it can prevent incorrect prior assumptions from driving the solution. 

The parameterised approach is especially useful in the context of exoplanet retrievals because the information content of data is continuously changing, and the complexity of parameterised models can very easily be tuned. For example, the early exoplanet retrieval model of \cite{madhu09} contained a six-parameter temperature-pressure profile, which effectively divided the atmosphere into three layers and described the temperature gradient within each layer. They also retrieved altitude-independent abundances of H$_2$O, CO$_2$, CO, CH$_4$ and NH$_3$, which were the five species they considered to be most likely to be active in the infrared. They found that they were unable to simultaneously fit the data from different instruments with the same model. A subsequent analysis by \cite{lee12} allowed the temperature to vary freely and smoothly as a function of pressure, which allowed a reasonable fit to be achieved to all datasets, but clearly included greater potential for model degeneracy due to the increased number of parameters. \cite{lee12} present correlations between the temperature-pressure profile and the abundances of the molecular species, demonstrating the extent of this degeneracy. 

Parameterisation has also evolved in modelling of primary transit spectra. The different geometries of primary and secondary transit observations mean that each is sensitive to different aspects of the atmospheric state, and so different parameters are included depending on the type of observation.

\begin{figure}
    \centering
    \includegraphics[width=0.95\textwidth]{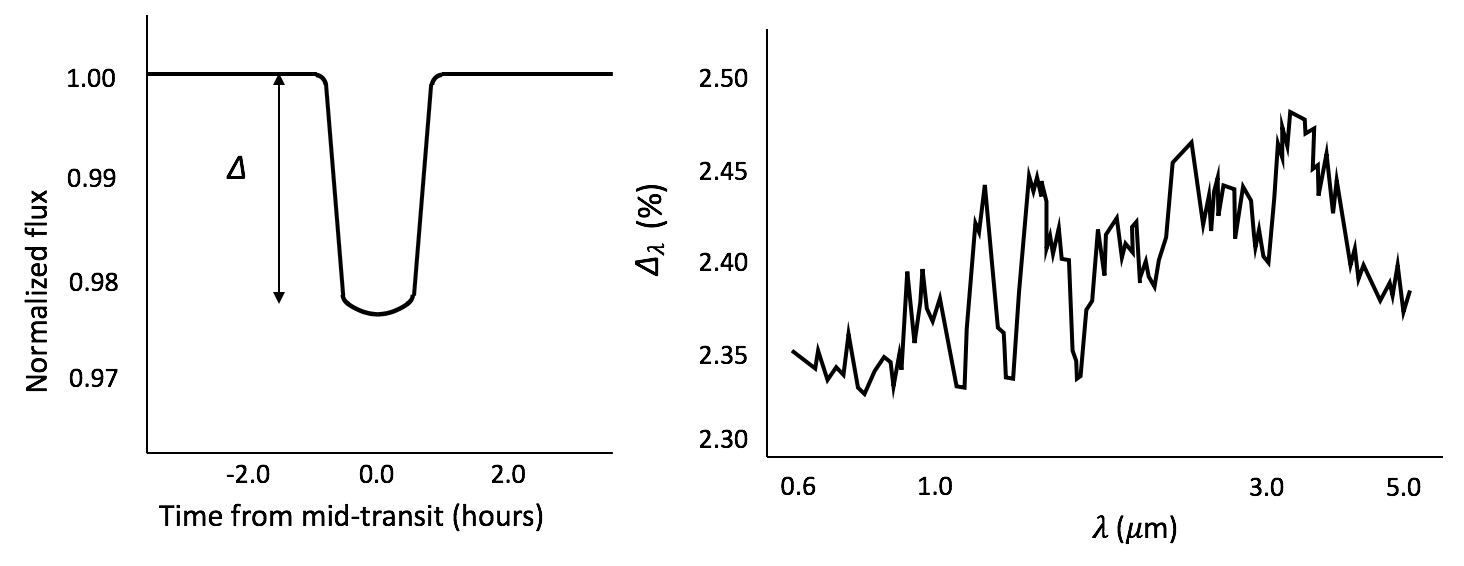}
    \caption{Sketches showing a transit lightcurve and the corresponding transit spectrum.}
    \label{transit_fig}
\end{figure}

The transit depth in primary transit is given by
\begin{equation}
  \mathrm{\Delta}_{\lambda} =  1 - \Big(\frac{R_{\mathrm{p,\lambda}}}{R_{\mathrm{s}}}\Big)^2
\end{equation}
where $R_{\mathrm{p,\lambda}}$ is the radius of the planet and $R_{\mathrm{s}}$ the radius of the star. A transit spectrum is the variation in transit depth as a function of wavelength, which results from the change in atmospheric opacity due to the presence of absorbing gases and aerosols (Figure~\ref{transit_fig}).

By contrast, the signal in secondary eclipse is obtained by measuring the difference in flux immediately before and after the eclipse, when the dayside of the planet is visible, with the flux of the star alone when the planet is in eclipse.

Primary transit observations do not probe deeper regions of the atmosphere, as the atmosphere becomes opaque to radiation passing tangentially through the atmosphere at lower pressures compared with radiation emerging close to nadir. Transit spectra also solely measure light from the star that has passed through the atmosphere rather than thermal emission from the planet itself; therefore, primary transit spectra are less sensitive than secondary eclipse spectra to the temperature-pressure profile. Retrievals covering only a small spectral range in primary transit have therefore generally assumed an isothermal temperature-pressure profile. By contrast, primary transit spectra are extremely sensitive to the atmospheric scale height $H$, as it is the physical thickness of the atmosphere that determines the amplitude of the spectral features in a primary transit observation.

The pressure $p(z)$ of an atmosphere in hydrostatic equilibrium may be defined as:
\begin{equation}
p(z) = p(0)e^\frac{-z}{H}
\end{equation}
where $p(0)$ is the surface pressure and $H$ is the atmospheric scale height.
\begin{equation}
H = \frac{kT}{{\mu}g}
\end{equation}
where $k$ is the Boltzmann constant, $T$ is temperature, $\mu$ is the mean molecular mass, and the local gravitational acceleration is
\begin{equation}
g = \frac{GM_P}{(R_P+z)^2}.
\end{equation}
Here, $G$ is the universal gravitational constant, $M_P$ is the mass of the planet, $R_P$ is the radius of the planet and $z$ is the altitude above the surface. The dependence of the scale height on $g$ means that there is significant sensitivity to the absolute radius of the planet (as opposed to the radius relative to that of the star). This emphasises the requirement for precise and accurate radii for planet host stars.

This in turn impacts what is variously referred to as the normalisation degeneracy or the baseline issue \citep{benneke12,griffith14,heng17}; because the planetary radius quoted in the literature is derived from the white light transit, the pressure that this represents is dependent on the atmospheric properties. Either the pressure at some given radius, or the radius at some given pressure, must therefore also be free parameters in the retrieval. Because the scale height is then proportional to both temperature and the square of the radius, these two quantities are degenerate and are inversely correlated in retrievals. \cite{fisher18} demonstrated that the normalisation degeneracy may be partially broken using low-resolution transmission spectra measured by \textit{Hubble}-WFC3 alone, because information on temperature and chemical abundances are encoded in the shape of the transmission spectrum. However, this degeneracy and others are more easily broken by including broad wavelength coverage data, as discussed in Section~\ref{sota_chem}.

Primary transit spectra are also affected by the presence of clouds. Effects can be dramatic to the point of cloud obscuring all molecular and atomic features in the spectrum (e.g. \citealt{kreidberg14}). In less extreme cases, the amplitudes of gas absorption features are reduced in the presence of cloud because the atmosphere becomes opaque below the cloud top, so only the centres of molecular bands are observed. This effect can be difficult to distinguish from either a) low abundances of the molecular species in question or b) a high mean molecular mass (and therefore low scale height) atmosphere.  

Simple 1D forward models for retrieval codes need to include parameterisations of these effects. Cloud is often treated as a completely grey, opaque layer with a variable top pressure (e.g. \citealt{kreidberg14}). This approach has the advantage of introducing only a single parameter, but is also not very representative of a real cloud, which is likely to have a wavelength-dependent optical depth and to be partially transparent at some wavelengths. We discuss cloud parameterisation in more detail in Section~\ref{sota_clouds}. The mean molecular mass may be specified as a separate free parameter, or may be calculated after the fact based on the retrieved abundances of the modelled gases; this approach is computationally simpler, but risks misinterpretation should large abundances of a spectrally inactive, heavy gas such as N$_2$ are present. It also relies on a complete range of molecular species being included in the model.

\section{Chemistry}
\label{chemistry}
In this section, we discuss the challenges of inferring information about chemistry, and thence planetary formation and origin scenarios, from the retrieved abundances of individual gases. We begin by summarising the current state of the art. Whilst there is a wealth of literature available dealing with detailed studies of individual planets, here we find it is more instructive to focus on works that analyse multiple planets, as this provides a more general indication of the degree to which atmospheric properties can be constrained with currently available data. 

\subsection{State of the art: chemistry}
\label{sota_chem}

Hot Jupiters observed in primary transit are ideal targets for molecular species detection and constraint, as these planets have large scale heights and therefore large feature amplitudes in primary transit (in the absence of clouds). Several comparative retrieval studies of hot Jupiters with \textit{Hubble Space Telescope} and \textit{Spitzer} observations have been recently performed, following on from the presentation by \cite{sing16} of near-infrared spectra of ten hot Jupiters with consistent data reduction. 

\textit{Hubble} Wide Field Camera 3 (WFC3) data are now available for several tens of exoplanets. Many of these also have photometry from the \textit{Spitzer} InfraRed Array Camera (IRAC) and spectra from the \textit{Hubble} Space Telescope Imaging Spectrograph (STIS). As WFC3 spectra are the most widely available, studies such as \cite{tsiaras18} and \cite{fisher18} focus on this dataset only. Because WFC3 has a relatively narrow wavelength range, between 0.8 and 1.6 $\upmu$m, only a subset of interesting molecular species can be constrained. The 1.4 $\upmu$m H$_2$O band dominates the spectral shape in this region, although features from TiO, VO and FeH may be discernable at the shorter wavelength end if present, along with CH$_4$, HCN and NH$_3$ longwards of 1 $\upmu$m. 

\cite{tsiaras18} use a 10-parameter model to study 30 hot and warm gaseous planets, including volume mixing ratios of H$_2$O, CO$_2$, CO, CH$_4$ and NH$_3$; isothermal temperature; planet radius; and three cloud parameters (discussed further in Section~\ref{sota_clouds}. For planets hotter than 1400 K they also include TiO and VO abundances. They define an atmospheric detectability index (ADI) which is the Bayes factor between the nominal atmospheric model and a straight line (featureless) spectrum, and they class any planet with ADI $>$3 as having a detectable atmosphere. They find that 16 of the 30 planets studied fulfil this criterion; H$_2$O is found to be present on all of these planets, with abundances typically constrained to $\pm$ an order of magnitude. No constraints are obtained for CO$_2$, CO, CH$_4$ or NH$_3$ on any planet, but for two (WASP-76b and WASP-121b) there is evidence that TiO and VO are present; a subsequent analysis including STIS data for WASP-121b by \cite{evans18} corroborates the presence of VO but not of TiO. 

\cite{fisher18} examine a similar dataset of 38 WFC3 transmission spectra, although their analysis extends to smaller and temperate planets, such as the warm mini-Neptune GJ 1214b and the likely rocky earth-sized planets TRAPPIST 1d--g. Unlike \cite{tsiaras18}, they only consider volume mixing ratios of H$_2$O, NH$_3$ and HCN in their model. They include a slightly more complex cloud parameterisation (see Section~\ref{sota_clouds}) and allow for a non-isothermal temperature profile. They retrieve a reference pressure rather than a reference radius for the planet. \cite{fisher18} find no evidence that the region of the atmosphere probed during transit deviates from an isothermal profile, and they conclude that most of these spectra may be explained by an isothermal transit chord containing only water and grey clouds.

Two further studies, \cite{barstow17} and \cite{pinhas19}, consider a smaller number of planets but take into account data from \textit{Hubble}/STIS and \textit{Spitzer}/IRAC. A broader wavelength range allows degeneracies between cloud properties and gas abundances to be broken, but this coverage is not available for as many planets, and the inclusion of spectral segments obtained at different times introduces the issue of stitching together non-contemporaneous spectra that may have been affected by instrumental and astrophysical systematics in different ways. For this reason, the datasets used are those provided by \cite{sing16}, in which spectra were consistently reduced in an attempt to minimise this issue. 

\cite{barstow17} uses a hybrid approach, combining the fast but prior-restricted optimal estimation retrieval method with a grid search to ensure exploration of a wide parameter space. Gases included in the retrieval are H$_2$O, CO$_2$, CO, CH$_4$, but there is no evidence for the presence of any gas except H$_2$O. Constraints on H$_2$O abundance are obtained for all planets except WASP-12b, which has poor quality WFC3 data in the \cite{sing16} paper, and WASP-6b and WASP-39b, for which no WFC3 data were available at the time. H$_2$O abundances are constrained only to within an order of magnitude, but show a clear trend towards subsolar abundances. This trend was also found by \cite{pinhas19}, who performed a nested sampling retrieval of the same dataset; \cite{pinhas19} used new WFC3 data for WASP-12b and WASP-39b, which allowed constraints on H$_2$O abundance for these planets also.  

Although \cite{barstow17} and \cite{pinhas19} use different cloud parameterisations, the H$_2$O abundance results are consistent with each other where the same data are used. The differing results for the cloud properties are discussed further in Section~\ref{sota_clouds}. 

\begin{figure}
    \centering
    \includegraphics[width=0.9\textwidth]{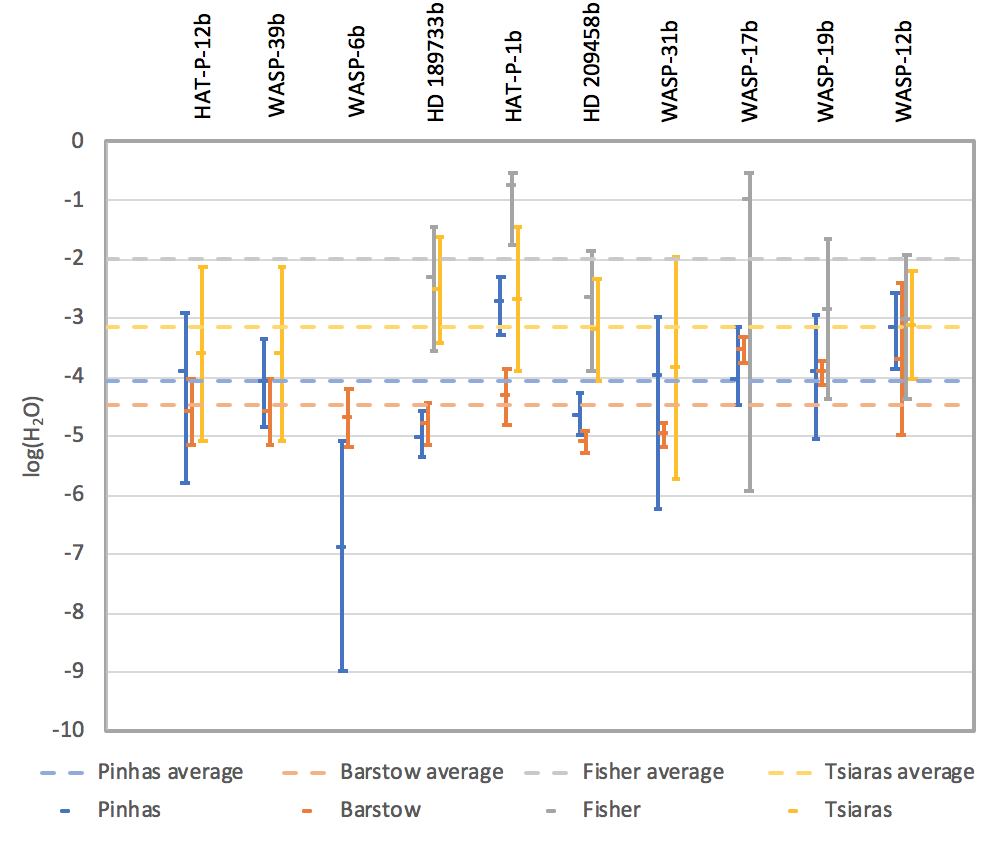}
    \caption{A comparison of H$_2$O volume mixing ratios retrieved using four different retrieval algorithms. \cite{pinhas19} and \cite{barstow17} use spectra combining \textit{Hubble}/STIS and WFC3, and \textit{Spitzer}/IRAC, whereas \cite{fisher18} and \cite{tsiaras18} use only \textit{Hubble}/WFC3. The dashed lines represent the error-weighted average abundances, excluding WASP-6b for which no WFC3 data is available.}  
    \label{h2o_comp}
\end{figure}

In Figure~\ref{h2o_comp}, we present a comparison of the retrieved H$_2$O abundances for each of the studies described above. The values shown for \cite{barstow17} are taken from the range of values from the best-fitting models for each planet; the central value shown is just the average of the minimum and maximum. All other values are obtained directly from the marginalised retrieval solution in each case. In general, retrievals accounting for \textit{Hubble}/STIS and \textit{Spitzer}/IRAC data converge on lower H$_2$O abundances, whereas solutions from just \textit{Hubble}/WFC3 have higher H$_2$O abundances. \cite{pinhas19} conclude that H$_2$O abundances are generally subsolar. Error-weighted averages are shown, calculated over all available planets except for WASP-6b, for which no WFC3 data are available. The very low H$_2$O volume mixing ratio retrieved for WASP-6b from \cite{pinhas19} is likely to be a result of a substantial drop in transit depth between the STIS spectrum and the the IRAC points in the infrared, which forces a scenario in which the spectrum is characterised by opaque haze and low gas abundances. Averages between \cite{barstow17} and \cite{fisher18} differ by more than two orders of magnitude, indicating that the inclusion of STIS and IRAC data is influential on the solution. The difference between the retrieval results with and without STIS and IRAC is most apparent for HD 189733b and HD 209458b. 

The likely reason for these differences when broader wavelength coverage data are added is that these data provide more constraints on cloud characteristics than WFC3 does by itself; muted H$_2$O features can either be the result of a low abundance of H$_2$O, or the presence of cloud. The detection of absorption features due to multiple gases can also break degeneracies between temperature and gas abundance, and low-amplitude features can also be a result of low temperatures. Following this logic, we expect to see substantial improvements with the launch of \textit{JWST}, which will provide extremely broad wavelength coverage (although it cannot cover the full spectral range simultaneously).

Primary transit observations are generally preferred for obtaining constraints on molecular species abundance, but \cite{line14} completed a comparative study of 9 hot Jupiters in emission and present retrieval results for molecular species. The spectral coverage and resolving power is extremely variable across the 9 objects, with HD 189733b having data from \textit{Hubble}/Near Infrared Camera and Multi-Object Spectrograph (NICMOS), \textit{Spitzer}/IRAC and \textit{Spitzer}/InfraRed Spectrograph (IRS), whereas the majority are restricted to photometric observations only. Good constraints on molecular abundances (beyond upper and lower limits) are generally only obtained for cases with spectroscopic data. In this case, volume mixing ratios for H$_2$O, CO$_2$ and CH$_4$ are constrained to within an order of magnitude for HD 189733b, and H$_2$O is similarly constrained for TRES-3b, but no further strong constraints are obtained for any of the other planets in the sample.

Subsequent publications looking at single planets in emission have obtained some constraints on molecular abundances. \cite{stevenson14} analyse the dayside spectrum of ultra-hot Jupiter WASP-12b. They test oxygen-rich (C:O $\sim$ 0.5) and carbon-rich (C:O $> \sim$ 1.0) atmospheric models, and find that the carbon-rich model is preferred, although their best-fit solution has what the authors consider to be implausibly high abundances of CH$_4$ and CO$_2$, and very low abundances of H$_2$O. An analysis of the same dataset by \cite{oreshenko17} shows that the solution is highly dependent on prior assumptions made about the chemistry. \cite{heng16} point out that it is nearly impossible to have CO$_2$ be more abundant than CO in H$_2$-dominated atmospheres unless the metallicity exceeds solar by about 3 orders of magnitude, and this constraint should be used to rule out chemically implausible retrieval solutions.  Some evidence for the presence of TiO and VO has also been reported from secondary eclipse observations, of WASP-33b (by \citealt{haynes15}) and WASP-121b (VO only, by \citealt{evans17}). 

Molecular abundance information from secondary eclipse spectra lags behind that available from transits, as secondary eclipse contrast improves at wavelengths beyond the reach of \textit{Hubble}, and the lack of cryogenic cooling for \textit{Spitzer} means that currently precise secondary eclipse spectra are hard to come by. This situation is expected to improve enormously once \textit{JWST} has launched. Despite significant advances in spectral quality for both primary transit and secondary eclipse over the last decade, precise abundance constraints are only reliably available for H$_2$O, and even this is not universally possible. The main barrier to molecular species constraint is the typically narrow wavelength range accessible for most planets; wavelengths beyond the red end of the \textit{Hubble}/WFC3 G141 grism ($>\sim$ 1.6 $\upmu$m) are required to constrain most molecular species apart from H$_2$O and metal oxides/hydrides, and spectral data in this range is currently unavailable. This situation will be vastly improved once \textit{JWST} has launched, as it will improve signal-to-noise and resolving power by at least a factor of 10, and push spectral coverage further into the infrared. Several predictive studies exist that indicate \textit{JWST} spectra will provide excellent opportunities for retrieval constraints on molecular abundances from both primary transit and secondary eclipse spectra of hot Jupiters (e.g. \citealt{barstow15,greene16}) and also allow the characterization of smaller, terrestrial worlds (e.g. \citealt{barstow16}, \citealt{krissansen-totton18}). 

\subsection{Recovery of underlying chemical trends}
\label{chemtrends}
A key part of the planetary formation/evolution puzzle is the bulk C:O ratio of a planet. It has been postulated that this is an indicator of where in the disc a planet has formed \citep{oberg11} as the location of the planet relative to the snowlines could affect the composition of the accreted material. \cite{oreshenko17} attempted this exercise for WASP-12b using an emission spectrum constructed from \textit{Hubble}-WFC3 and \textit{Spitzer}-IRAC, and suggested that WASP-12b experienced disk-free migration during its formation history. Determining the bulk C:O ratio from spectroscopy has already been attempted in exoplanet retrievals (e.g. \citealt{line14,kreidberg15}), although so far this is hampered by a lack of access to regions of the spectrum containing features of carbon species. Observations by \textit{JWST} will alleviate this aspect of the problem, but the question remains to what degree of precision underlying chemical trends such as the C:O ratio can be recovered. This is particularly important in the context of future missions such as \textit{ARIEL}, which aims to provide the first exoplanet atmosphere population study. 

\cite{kreidberg15} compare retrievals with free chemistry (where each gas is retrieved individually) and retrievals of metallicity and C:O ratio under the assumption of equilibrium chemistry for the \textit{Hubble}/WFC3 spectrum of WASP-12b. The results for each case are in agreement in terms of the retrieved temperature and H$_2$O abundance, where H$_2$O is the only gas that can be constrained. Based on the assumptions within the chemical equilibrium model, \cite{kreidberg15} reject a carbon-rich atmosphere scenario at $>3\sigma$ confidence, as the retrieved H$_2$O abundance is higher than predicted for a carbon-rich model. However, this result is dependent on the assumptions within the chemical model used, so is somewhat less agnostic than a free-chemistry retrieval would be; there is a trade off between obtaining a tighter constraint and relying on a potentially flawed chemical model.

The only way to reliably demonstrate recoverability of underlying chemical trends is to conduct blind tests of retrieval algorithms on synthetic observations with known chemistry. There are two distinct facets to this challenge; 1) can the correct atmospheric C:O ratio be recovered for the constituents present within the observable atmosphere of the planet? and 2) can the correct planet bulk C:O ratio be recovered from the atmospheric C:O? The first issue simply relies on the ability of a retrieval algorithm to accurately determine the abundances of molecular and atomic species within a planet's atmosphere, whereas the second encompasses scenarios in which the bulk planet chemistry is not reflected in the molecular make up of the atmosphere, for example because substantial amounts of some elements are present in the form of clouds deep in the atmosphere. An illustration of this difficulty is the challenge of determining the H$_2$O volume mixing ratio in Jupiter's atmosphere; see e.g. \cite{li20}. Simple tests can be performed to answer question 1) with 1D forward models containing some parameterised chemistry, but for question 2) more complex models following through from planet formation to the eventual atmospheric composition will be required. 

In the short term, studies testing the ability to accurately recover chemical trends in atmospheric composition should be undertaken. Efforts in this direction are already underway in preparation for the \textit{ARIEL} mission, but similar studies are required for other datasets as the information content of spectra is highly dependent on the precise details of resolving power and wavelength coverage. 

\textbf{Recommended action: conduct retrievals of simulated datasets with known atmospheric chemistry, for a range of planetary temperatures and metallicities, as observed by a variety of instruments. This will allow us to determine observational requirements for precise constraints on C:O ratio, and other trends of interest e.g. N:O ratio.}

\section{Temperature structure}
\label{temp}
Whilst detailed information about temperature structure is difficult to obtain from primary transit observations due to the relatively narrow pressure range that is probed, temperature-pressure profiles have been retrieved from secondary eclipse and phase curve spectra. Whilst very broad spectral coverage, such as that available for HD 189733b (e.g. \citealt{lee12,line14}), probes a sufficient range of atmospheric pressures to allow a smoothed, free retrieval of temperature as a function of pressure, the majority of secondary eclipse spectra cover a smaller range and parameterisation is necessary to extrapolate the atmospheric structure beyond the region that is directly constrained.

\subsection{State of the art: temperature-pressure profiles}
\label{sota_temp}
The simplest approach to retrieving temperature is to make the crude assumption that the temperature profile is isothermal. This has often been the approach taken when analysing primary transit spectra; however, \cite{rocchetto16} show in their synthetic retrieval study for the \textit{James Webb Space Telescope} that this assumption can result in errors of more than an order of magnitude in the retrieved gas abundances for some cases. The isothermal approximation is therefore clearly inadequate, and approaches that capture the broad shape of the temperature structure must be explored. 

There are two parameterisation approaches favoured by retrieval groups, the simpler of the two being the Guillot profile \citep{guillot10} which has 5 free parameters and was first implemented by \cite{line12}, and the other being the approach advocated by \citet{madhu09}, which we will call the Madhusudhan profile, and has 6 free parameters.  The original Guillot profile assumes that no scattering occurs in the atmosphere; \cite{heng12} and \cite{heng14} respectively generalised the Guillot profile to include isotropic scattering (by either aerosols or atoms and molecules), and non-isotropic scattering (large particles).

The Guillot profile is based on a three-channel approximation for an atmosphere in thermal equilibrium and is described by the following equation,
\begin{equation}
T^4(\tau) = \frac{3T_{\mathrm{int}}^4}{4}\left(\frac{2}{3}+\tau\right) + \frac{3T_{\mathrm{irr}}^4}{4}(1-\alpha)\xi_{\gamma_{1}}(\tau) + \frac{3T_{\mathrm{irr}}^4}{4}(\alpha)\xi_{\gamma_{2}}(\tau)
\end{equation}
where 
\begin{equation}
\xi_{\gamma_i} = \frac{2}{3} + \frac{2}{3\gamma_i} \left[ 1 + \left( \frac{\gamma_i \tau}{2} - 1 \right) e^{-\gamma_i \tau} \right] + \frac{2\gamma_i}{3} \left( 1- \frac{\tau^2}{2}\right) \mathrm{E}_2(\gamma_i\tau)
\end{equation}
and the irradiation temperature is
\begin{equation}
\label{eqn4}
T_{\mathrm{irr}} = \beta \left( \frac{R_{\star}}{2a} \right)^{1/2} T_{\star}
\end{equation}

The 5 free parameters are $\kappa_{\mathrm{IR}}$, the infrared opacity; $\gamma_1$ = $\kappa_{\mathrm{v1}}/\kappa_{\mathrm{IR}}$ and $\gamma_2$ = $\kappa_{\mathrm{v2}}/\kappa_{\mathrm{IR}}$, the ratio of two-band visible opacities to the IR opacity; $\alpha$, the ratio of the flux between the two visible streams; and $\beta$ is a measure of the recirculation efficiency of the atmosphere. $T_{\mathrm{int}}$ is the planet's internal temperature, and $T_{\mathrm{irr}}$ is the temperature calculated from irradiation by the parent star. $R_{\star}$ and $T_{\star}$ are the radius and temperature of the parent star, and $a$ is the orbital semi-major axis. $\tau$ = $\kappa_{\mathrm{IR}}p/g$ is the infrared optical depth of the atmosphere, where $p$ is atmospheric pressure and $g$ is gravitational acceleration. E$_2$ is the second order exponential integral function. 

The Madhusudhan profile divides the atmosphere into three layers. Layer 1 is the uppermost and is bounded at the base by pressure $P_1$. Layer 2 extends from pressure $P_1$ to $P_3$, and Layer 3 extends downwards from $P_3$. The temperature in each layer is defined as follows:

$P_0 < P < P_1~~~P=P_0\mathrm{e}^{\alpha_1(T-T_0)^{\beta_1}}$

$P_1 < P < P_3~~~P=P_2\mathrm{e}^{\alpha_2(T-T_2)^{\beta_2}}$

$P > \textit{P}_3~~~T = T_3$

In all cases, $P_0 < P_1 < P_3$. If the temperature profile is inverted, $P_1 < P_2 < P_3$; if not, $P_1 \ge P_2$.  This can be simplified to only 6 free parameters by setting $P_0$ equal to the pressure at the top of the atmosphere; empirically setting $\beta_1$ = $\beta_2$ = 0.5. Finally, the temperature profile is forced to be continuous at the boundaries between the layers where $P$ = $P_1$ and $P$ = $P_3$, which leaves 6 free parameters: $P_1$, $P_2$, $P_3$, $\alpha_1$, $\alpha_2$ and $T_3$. 

The advantage of the Guillot profile is that the shape is physically motivated by the assumption of radiative equilibrium, whilst still being a fairly simple parameterisation. It does however contain a bias in that it produces isothermal profiles at low pressures, which may not be an accurate reflection of a real atmosphere. This isothermal behavior is a subtle artefact of using mean opacities, where ``mean" in this case is ill-defined.  Specifically, in order for the solution to be analytically tractable, the derivation assumes that the absorption, flux and Planck mean opacities are equal. The Madhusudhan profile allows more flexibility of shape, particularly with regards to resolving temperature inversions, at the expense of an additional free parameter. \cite{blecic17} investigate the ability of such 1D temperature parameterisations to recover the temperature structure from synthetic eclipse spectra generated from 3D atmospheric circulation models. They find that the Madhusudhan profile provides a better match to the temperature structure in the middle atmosphere as it is more capable of producing an inversion; however, it does not match the deep temperature structure. We discuss the reliability of fitting a 1D temperature model to a dataset generated from a 3D circulation model in Section~\ref{sota_3d}. 

A key science question relating to T-p profile retrievals is the presence or absence of a temperature inversion in hot Jupiter atmospheres. Inversions were predicted to occur in planets with incident flux of greater than 10$^9$ erg s$^{-1}$ cm$^{-2}$, due to the presence of optical absorbers TiO and VO in their atmospheres \citep{fortney08}. This category includes several well-studied hot Jupiters such as HD 209458b, but so far only a handful of planets show evidence for thermal inversions in their dayside spectra. These include WASP-33b (\citealt{haynes15}; fit using Madhusudhan profile); WASP-121b (\citealt{evans17}; fit using Guillot profile); and WASP-18b (\citealt{sheppard17}; fit using Madhusudhan profile). All of these are ultra-hot Jupiters with equilibrium temperatures of over 2000 K, suggesting that the cut-off irradiation for thermal inversions is somewhat higher than originally predicted. 

Reliably retrieving the dayside temperature structure is further complicated by the presence of solution degeneracy with gas abundance retrievals. \cite{stevenson14} retrieve the dayside atmospheric state for WASP-12b and test two models which force either carbon-rich or oxygen-rich chemistry; the retrieved temperature profiles differ by several hundred K at low pressures. Similarly, \cite{barstow14} test the effect of varying gas abundance priors on a continuous Optimal Estimation retrieval of temperature from HD 189733b emission spectra, and find that the precise shape of the profile is dependent on the gas abundance prior chosen. 


\subsection{Future challenges for temperature parameterisation}
\label{future_temp}

Investigations are underway into the most appropriate temperature parameterisations in the \textit{JWST} era and beyond. \cite{rocchetto16} simulate several \textit{JWST} hot Jupiter transmission spectra for model atmospheres with varying C:O ratios, and they demonstrate that oversimplified parameterisations in temperature structure retrieval can introduce significant bias in other retrieved properties. Assuming that the temperature profile is isothermal can result in, for example, retrieved CO abundances over an order of magnitude too high. A Guillot temperature-pressure profile, whilst it increases the uncertainty on the retrieved properties, results in a more accurate retrieval of the gas abundances. However, it is important to note that the input temperature-pressure profile is close to the typical shape predicted by the Guillot parameterisation, so the ability of the Guillot profile to achieve a good fit may be serendipitous. It is clear, therefore, that accurate chemistry retrievals are dependent on the suitability of the temperature parameterisation.

This issue is likely to only become more complex as the information content of the spectrum increases. The key difficulty in transmission will still be the relatively small pressure range (when compared with eclipse spectra) probed by the observation, and the degeneracy between the effects on temperature, mean molecular weight and gravity on the scale height. Further investigations of the kind presented by \cite{rocchetto16} are likely to be a critical aspect of model development. Ultimately, the ideal for eclipse spectra would be to explicitly retrieve temperature at each level in the model atmosphere, subject to some correlation length to ensure smoothness, but this is likely to only be possible for the very highest signal-to-noise observations. 

\textbf{Recommendation: conduct retrievals of simulated datasets with a variety of temperature structures and chemistry, to investigate regions of parameter space where the temperature profile parameterisation introduces most bias. Investigate alternative approaches to those currently in the literature for intractable cases.}

\section{Clouds}
Initial attempts to characterise the atmospheres of hot exoplanets via retrieval were conducted without reference to clouds, due to the erroneous belief that the extreme temperatures would make it impossible for clouds to exist. The inclusion of clouds also inevitably complicates the retrieval process, as it introduces further parameters into what is already an underconstrained retrieval problem. Clouds are complex, potentially spatially variable, structures that provide broadband absorption and scattering, and as such affect spectra in ways that can be difficult to identify. They can also have the effect of muting molecular absorption features. 

\subsection{State of the art: clouds}
\label{sota_clouds}
So far, retrieval efforts have used simple parameterisations to try and capture the cloud properties that produce the most significant effects on spectra. The different geometries of exoplanet observations require different treatment; in primary transit, due to the long path length through the atmosphere what is often referred to as the cloud top pressure is especially important because the atmospheric opacity rapidly increases below the cloud top. Conversely, the cloud top pressure is less critical if the planet is being directly imaged in the infrared, as the measured radiation is emerging from the planet beneath the cloud top. 

\begin{figure}
    \centering
    \includegraphics[width=0.9\textwidth]{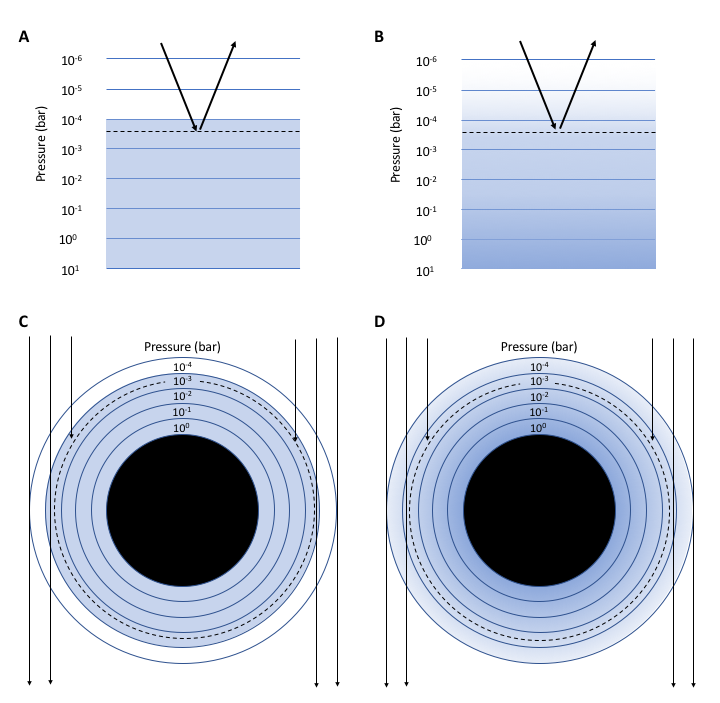}
    \caption{Effective cloud top pressure for different models in both reflection (nadir; panels A and B) and transit (limb; panels C and D) geometry. In A and C, the cloud has a uniform specific density below a cloud top pressure of 10$^{-4}$ bar; in B and D, the cloud has a specific density that decreases with decreasing pressure, with the cloud extended throughout the atmosphere. The level in the cloud at which the optical depth is unity (dashed line) is the same for both cloud models in each geometry, even though the vertical distribution of aerosol is very different.}
    \label{cloud_schematic}
\end{figure}

Cloud top pressure is not in reality a well-defined pressure above which cloud ceases to exist, although it can be treated as such in simple parameterisations. It represents the pressure level at which the cloud optical depth is unity, which is highly dependent on the observation geometry - the cloud optical depth reaches unity at a higher altitude in limb geometry compared with nadir. The effective cloud top pressure can be altered in a simple model by setting a physical cloud top, or by varying the opacity of a cloud that is not confined to any particular pressure range. These two approaches are not exactly equivalent, so two different models with different predicted spectra could have the same effective cloud top pressure (Figure~\ref{cloud_schematic}).

The simplest primary transit studies have assumed that the atmosphere is completely opaque at all wavelengths, for pressures higher than the cloud top pressure. This is suitable over relatively small wavelength ranges, and for planets with spectra that are flat over a wide wavelength range (e.g. GJ 1214b, \citealt{kreidberg14}; \citealt{fisher18}). However, in general this would only be representative of a cloud made of large particles with a broad size distribution, and fails to account for scenarios where aerosols may more closely resemble small-particle haze. Slightly more complex parameterisations allow for the possibility of optically thin clouds, and a simple power law for extinction as a function of wavelength (e.g. \citealt{barstow17,pinhas19}.) 

Whilst the pressure at the cloud top and the extinction slope are the most important parameters for primary transit, the vertical distribution of the cloud below the cloud top may also be important, depending on the cloud optical thickness. It is also possible that the cloud consists of multiple components - for example, an optically thin, small particle haze layer overlying an optically thick cloud \citep{macdonald17,pinhas19}. This has led to a range of different parameterisation options even just within primary transit retrievals, which can produce different and apparently contradictory results when applied to the same dataset. For example, retrievals of the same HD 189733b dataset by \cite{barstow17} and \cite{pinhas19} give consistent values for the H$_2$O abundance of $\sim$10$^{-5}$, but the retrieved cloud properties appear dramatically different at first glance. \cite{barstow17} characterize the HD 189733b cloud layer as a vertically thin, high Rayleigh scattering haze layer, whilst \cite{pinhas19} retrieve a cloud top deep in the atmosphere. However, this retrieved cloud top is the top of an opaque, grey cloud, which is coupled to a scattering haze layer for $P<P_{\mathrm{top}}$. Therefore, results from both parameterisations are in agreement that there is no visible grey cloud layer, and are consistent with the presence of scattering, small particle haze higher in the atmosphere. 

More complex parameterisations that include some information about composition have also been tested. \cite{kitzmann18} develop a parameterisation based on analytical fits to expected extinction cross-section curves of potential cloud species, such as e.g. MgSiO$_3$. The extinction efficiency $\kappa$ as a function of wavelength is parameterised as follows:

\begin{equation}
    \kappa_{\mathrm{cloud}} = \frac{\kappa_0}{Q_0 x^{-a} + x^{0.2}}
\end{equation}

where $\kappa_0$ is a scaling factor, $Q_0$ determines the wavelength at which the extinction efficiency peaks and is related to the cloud composition, $a$ is a scattering slope index and $x$ is the particle size parameter, given by

\begin{equation}
    x = \frac{2{\pi}r}{\lambda}
\end{equation}

This parameterisation is more easily related to real physical characteristics of cloud, such as particle size and composition. So far, it has been applied to \textit{Hubble}/WFC3 data by \cite{fisher18}, which provides relatively little constraint on cloud properties; it has not yet been applied to data spanning a broader wavelength range. 

The limited information available from current spectra, and the range of possible ways cloud can be represented, makes interpretation of these retrievals very difficult. Without a good understanding of the precise effects of different parameterisations on the spectrum, erroneous conclusions can be drawn. 

Attempts have also been made to consider cloud for secondary transit and directly imaged spectra. \cite{barstow14} consider the effect of clouds on the HD 189733b reflection spectrum observed by \cite{evans13}, but due to the requirement to include multiple scattering for reflection spectra only a simple grid search was performed. The cloud properties showed substantial degeneracy with the sodium abundance in the visible part of the spectrum. 


\subsection{Future challenges for clouds}
\label{future_clouds}
\label{cloudtemp}
Current exoplanet retrieval efforts are already demonstrating that the details of parameterisation for cloud properties have the potential to bias results. In the case of cloud properties, gas abundance retrievals seem to be somewhat immune to the differences in cloud treatment, but the conclusions drawn about the clouds themselves can vary widely, as discussed in Section~\ref{sota_clouds}. The main challenge we face here is to tune complexity of parameterisation to the information content of the data, whilst avoiding where possible introducing bias into the retrieval. Again, the only way to guard against this is to conduct rigorous simulation tests of retrieval parameterisations. 

Recent work has been undertaken to combine cloud microphysics models with 3D circulation models, and to use this to predict emergent spectra \citep{lines18}. Whilst we do no expect these simulations to perfectly predict real cloud and haze in exoplanet atmospheres, the ability of the retrieval scheme to recover key parameters from these synthetic spectra is an important test of the cloud parameterisation used. It provides an opportunity to check whether the parameterisation is sufficient to represent the spectral effect of complex cloud structure, and ensure that it does not introduce bias into the retrieval. Several different approaches to modelling cloud microphysics (e.g. \citealt{helling08,ackerman01}) and including cloud in GCMs (e.g. \citealt{lee17,parmentier16,mendonca18}) are available; the ideal would be a parameterised model that can recover key cloud properties from this range of available cloud models, whilst also accurately retrieving other atmospheric properties. 

\textbf{Recommended action: conduct retrievals of simulated datasets based on more detailed, physically motivated, 3D cloudy atmosphere models. Test a variety of simple cloud parameterisations, for a range of observational geometries, and compare results.}

\section{Phase curves and 3D effects}
For a handful of the most favourable targets, spectroscopic phase curves have been obtained which have allowed phase-resolved retrievals to be undertaken. The first example of this is the \cite{stevenson14b} phase curve retrieval for WASP-43b, obtained using \textit{Hubble}/WFC3. The limited wavelength coverage means there is only sensitivity to temperature structure over a small pressure range, and some information about the H$_2$O abundance. Difficulties of interpretation are compounded because the pressure of weighting function peak varies with phase, so comparison between phases is not straightforward. Phase curve observations with broader spectral coverage would resolve these difficulties and are planned for \textit{JWST}.  \cite{mendonca18} re-analyzed the Spitzer data of WASP-43b and ran cloudy GCMs to jointly analyze the Hubble and Spitzer phase-resolved emission spectra.  They find that the dayside is consistent with being cloudfree, with clouds confined to the nightside, and tentative evidence for elevated levels of carbon dioxide.

\begin{figure}
    \centering
    \includegraphics[width=0.95\textwidth]{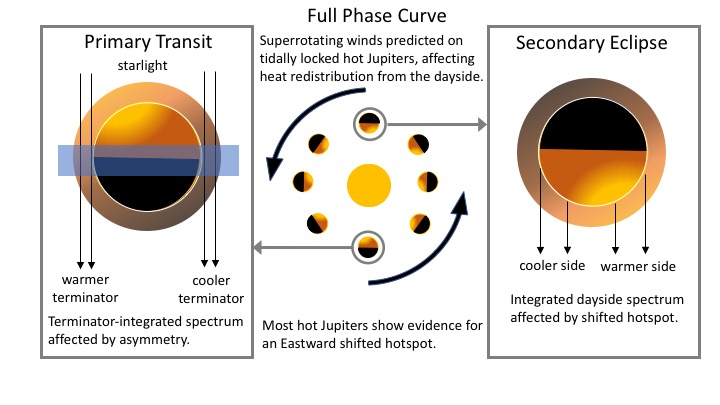}
    \caption{Strong superrotation on hot Jupiters, coupled with extreme irradiation, results in significant variation in temperature around the terminator region as observed in transit, and an asymmetric pattern on variation on the dayside as observed in eclipse.}
    \label{3D_schematic}
\end{figure}

Whilst phase curves provide some direct information about spatial variations in the thermal emission from the planet (and in some cases the reflected light), spatial variation in the atmospheric properties can also affect transmission spectra, albeit in a more subtle way. Evidence from observed phase curves and GCMs suggests that one terminator is likely to be hotter than the other for hot Jupiters, which will in turn impact the terminator chemistry and cloud coverage. For example, \cite{mendonca18} ran GCMs with disequilibrium chemistry (using a method known as ``chemical relaxation") and demonstrated that the coupling between atmospheric dynamics and chemistry produces spatial inhomogeneities across latitude, longitude and pressure for molecules such as water, and cannot be neglected if one wishes to accurately model phase-resolved spectra or wavelength-dependent phase curves.  The challenge in interpretation is that transmission spectra are averaged over the whole terminator region, so observations are implicitly 1D. Similarly, for planets not sufficiently favourable for us to have phase curve observations, secondary eclipse spectra are also 1D integrations over a non-uniform (and asymmetric) disc; the structure and chemistry retrieved using a 1D model will represent some sort of disc average, but it is unclear exactly what this corresponds to (Figure~\ref{3D_schematic}).

Likewise, the transit spectroscopy technique relies on the stellar disc being uniform once stellar limb darkening is corrected for, since it makes the implicit assumption that the planet transits a region of the stellar disc that is representative of the whole. This is of course not the case; stellar surfaces are highly non-uniform, with time-variable coverage of features such as spots and faculae. Spots and faculae have different spectral characteristics compared with the rest of the stellar disc, so unknown spot/faculae coverage fractions could lead to misinterpretation of transit spectra (e.g. \citealt{rackham18}). 

Directly imaged planets are relatively free of these issues, since they are not highly irradiated and their observations do not depend on the uniformity of the star's behaviour; we expect them to more closely resemble the Solar System giant planets in terms of their dynamics. However, we cannot rule out spatial asymmetry on these objects; whilst the dynamical regimes of hot Jupiters result in strong longitudinal gradients in temperature, the Solar System giants display latitudinal variation in chemistry and cloud properties (see e.g. PH$_3$ abundance on Jupiter and Saturn, \citealt{fletcher09}; in addition to the equator-pole differences observed on Jupiter, Saturn also has strong north-south seasonal asymmetry due to its axial tilt of 26.7$^{\circ}$). 

\subsection{State of the art: 3D effects}
\label{sota_3d}
\subsubsection{Phase curve retrievals}
For WASP-43b, a moderately hot Jupiter, it has been possible to obtain a spectroscopic phase curve using the \textit{Hubble}/WFC3 instrument. This allows retrievals to be performed as a function of phase, allowing longitudinal variations in chemistry and temperature structure to be mapped. \cite{stevenson14b} use the CHIMERA retrieval algorithm to analyse temperature structure at 16 different phases. The model includes 6 molecular absorbers, but only H$_2$O has a significant influence on the spectral characteristics. The temperature structure is modelled using 5 free parameters, after the method presented by \cite{parmentier14}. The retrieved upper atmosphere temperatures vary by 1000 K between the dayside and nightside, implying inefficient recirculation. 

So far, this is the only planet for which a full spectroscopic phase curve exists, so further exploration of phase curve retrievals is hindered by a lack of available data. Retrieval algorithms have not been applied to single- and multi-channel photometric phase curves that exist for other planets, presumably because the problem would be highly degenerate. However, spectroscopic phase curve observations are likely to be a priority for \textit{JWST}. WASP-43b is particularly well-suited to such observations as it has a very short period of only 19.52 hours; this planet will be re-observed at longer wavelengths with the Mid-InfraRed Instrument (MIRI) during the \textit{JWST} Early Release Science programme \citep{batalha17}, which will provide stronger constraints on the variation in atmospheric properties with phase. A phase curve for WASP-43b will also be obtained with the shorter wavelength NIRSpec instrument as part of the Guaranteed Time Observation for the instrument team \citep{birkmann17}.  

\subsubsection{3D cloud effects in transmission}
Work is already in progress to account for terminator asymmetry in retrieval models (e.g. \citealt{line16,macdonald17}), although so far it is restricted to cloud coverage, which ignores the fact that temperature structure, and likely the chemistry too, will also vary. \cite{line16} demonstrate that, over narrow wavelength ranges such as those probed by \textit{Hubble}/WFC3 only, partial terminator cloud cover is degenerate with cloud-free, high mean molecular weight atmosphere scenarios. Over a wider wavelength range, this degeneracy can be broken. \cite{pinhas19} include terminator cloud fraction in their retrieval of \textit{Hubble}/STIS + WFC + \textit{Spitzer}/IRAC spectra, and they recover a range of values between $\sim$0.2 and $\sim$0.8 for the 10 planets in their sample. They find no correlation between cloud fraction and any other key parameters in the study. \cite{line16} analyse WFC3 data only for HD 189733b, and find a cloud fraction that is comparable with the result from \cite{pinhas19}. 

\subsubsection{3D temperature structure from eclipse spectra}
\begin{figure}
    \centering
    \includegraphics[width=0.95\textwidth]{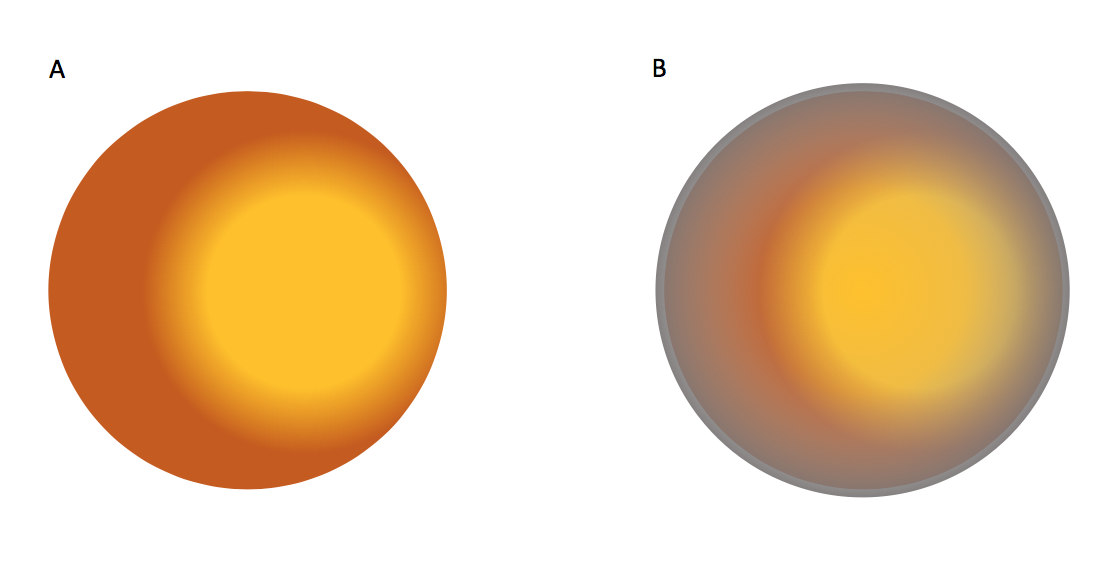}
    \caption{Panel A illustrates the contributions from each part of the dayside disc where the emission angle is not taken into account, whereas panel B shows which parts of the planet would dominate the signal when emission angle is accounted for.}
    \label{3D_weighted_mean_schematic}
\end{figure}

\cite{blecic17} investigate the ability of a 1D parameterised model to recover an average temperature structure from a simulated dayside spectrum generated from a 3D model atmosphere. They test both the Guillot and Madhusudhan temperature parameterisations discussed previously in Section~\ref{sota_temp}. Both parameterisations produce a retrieved temperature profile close to the arithmetic mean of the circulation model temperature profiles across the dayside, which is somewhat odd; the amount of radiation detected from different regions of the dayside is weighted by the cosines of the latitude and longitude, so it should follow that the hemisphere-integrated temperature-pressure profile should be a weighted average rather than a straightforward arithmetic mean (Figure~\ref{3D_weighted_mean_schematic}). This is an indication that further development of 1D retrieval models, and an investigation into surprising results such as this one, are required to reliably interpret these hemispherically averaged spectra. 

\cite{feng16} test the impact of using two temperature-pressure profiles to represent the hotter/colder regions of a planetary disc. They apply this to the first-quarter observation of WASP-43b, for which the visible portion of the planet is half in daylight and half in shadow, maximising the expected contrast. They also test simulated spectra for both current state-of-the-art observational scenarios (\textit{Hubble}+\textit{Spitzer}) and future observations with \textit{JWST}. They find that there is insufficient evidence with current data to favour a more complex model, but that for \textit{JWST} simulations significant biases in gas abundances are introduced when only a single temperature-pressure profile is used to represent the temperature structure. This approach is shown to work well where the temperature variation is adequately represented by two temperature-pressure profiles of equal weight, but it remains to be seen whether this is appropriate in the context of secondary eclipse, where the hotspot is likely to dominate. 

\subsubsection{Stellar heterogeneity in retrievals}

Parameterisation of the effects of starspots and faculae is now starting to be included within exoplanet retrieval frameworks. Initial results for super Earth GJ 1214b are presented by \cite{rackham17}. \textit{Magellan} telescope observations are fit using the CPAT absorber model coupled with a Markov-Chain Monte Carlo algorithm. The CPAT model for describing stellar heterogeneity divides the stellar disc into occulted and unocculted fractions. The wavelength-dependent transit depth, instead of being simply given by 
\begin{equation}
  \mathrm{\Delta}_{\lambda} =  1 - \Big(\frac{R_{\mathrm{p,\lambda}}}{R_{\mathrm{s}}}\Big)^2
\end{equation}
where $R_{\mathrm{p,\lambda}}$ is the radius of the planet and $R_{\mathrm{s}}$ the radius of the star, is instead given by
\begin{equation}
  \mathrm{\Delta}_{\lambda} = 1- \frac{(R_{\mathrm{p,\lambda}}/R_{\mathrm{s}})^2S_{\mathrm{o}}}{(1-F)S_{\mathrm{o}} + FS_{\mathrm{u}}}
\end{equation}
where $S_{\mathrm{o}}$ is the spectrum of the star in the occulted region, $S_{\mathrm{u}}$ is the spectrum of the star in the unocculted region, and $F$ is the fraction of the disc that is unocculted. \cite{rackham17} test PHOENIX \citep{husser13} model spectra with different metallicities, and different temperatures  as a proxy for varying levels of absorption across the stellar disc. They retrieve metallicity/temperature contrast between the occulted and unocculted regions and a constant offset in $R_{\mathrm{p}}$/$R_{\mathrm{s}}$, finding that the observed optical spectrum can be described by a case where 3.2 \% of the unocculted disc is 350 K hotter than the rest of the disc. This may be explained by starspot or facula contrast.

\cite{pinhas18} perform a retrieval analysis of nine hot Jupiters (the sample from \cite{sing16} excluding HD 189733b) using the same model as that presented in \cite{pinhas19} but also including stellar hetereogeneity. This is parameterised by the temperature of the heterogeneous regions (with the star's measured average photospheric temperature fixed) and the fractional coverage of any heterogeneities. \cite{pinhas18} do not discuss these values in detail, but instead present the model evidence for inclusion of stellar effects. They find substantial evidence of stellar heterogeneity for WASP-6 and WASP-39; whilst WASP-6 is one of the two most active stars in the sample based on log$R_{\mathrm{HK}}$ index, WASP-39 is less active, and for the most active star (WASP-19) the evidence is substantially against there being any stellar heterogeneity. This would indicate that the log$R_{\mathrm{HK}}$ index is an unreliable estimator of the importance of stellar heterogeneity effects on transit spectra.

\subsection{Future challenges for recovering planetary and stellar spatial information}
Two key resources for exploration of our ability to recover 3D information about planets are Global Circulation Models (GCMs; e.g. \citealt{selsis11,rauscher12,charnay15,amundsen16,lee16,parmentier16,mendonca18}), and the Solar System planets. GCMs are based on our current best understanding of the physical processes on hot Jupiters and young directly-imaged planets, and should be able to predict the broad characteristics of spatial variability on these planets. However, there are limits to the predictive power of GCMs due to the inability to accurately specify and represent all sources of dissipation in the atmosphere, e.g, \cite{goodman09,heng11,fromang16}; on Earth, these uncertainties can be mitigated by empirically calibrating the sources of dissipation in the GCM using in-situ data, an approach that is impossible for exoplanets.

The Solar System giant planets on the other hand, whilst they exist in a very different temperature/dynamics regime to the majority of well-studied exoplanets, have the advantage that we can directly compare spatially resolved datasets with the information that we would be able to recover if the planet was treated as a point source. This allows us to investigate for real objects how much information about large scale atmospheric ability and asymmetry persists in disc-integrated observations.

Models will also be key for understanding the impact of stellar heterogeneities on transmission spectra. As shown by \cite{rackham17}, whilst monitoring of target stars can provide an indication of the amplitude of variation in spot coverage, this does not provide information about the baseline level or the relative contributions of spots and faculae, both of which are important for transmission spectra. Understanding typical distributions and sizes of spots/faculae on different types of star will be extremely important for future observations. 

\textbf{Recommended actions: use simulated datasets from GCMs/stellar atmosphere models to test the ability of parameterised retrieval models to recover 3D information about the planet and the star. Investigate how the information content of spatially resolved observations of Solar System giants compares with that of the same observation degraded to a point source.}

\section{Conclusions}
We have presented a summary of the current and imminent future challenges surrounding atmospheric retrievals of exoplanets. In general, the obstacles faced result from the lack of available ground truth for exoplanet observations, and, especially in the near future, a rapid increase in the information content of observations which requires modelling strategies to constantly evolve. 

A common theme for solutions to these challenges is the use of physically based climate and circulation models to provide simulated datasets. Whilst we cannot yet be sure that the outputs from these models are accurate representations of real exoplanet atmospheres, they do allow us to perform important tests of how well simple parameterised models capture more complex atmospheric characteristics. In the case of stellar heterogeneity, it is likely that we will have to rely to some extent on ab initio stellar atmosphere models if we want to correct for spectral contamination of starspots and faculae. 

Another key attribute required for retrieval models is flexibility; since the data quality is, and is likely to remain, variable across different planets, it is important that models can be easily tuned to maximally exploit the information content of a given observation. Oversimplification has been demonstrated to introduce bias - for example, assuming an isothermal temperature structure for broad wavelength coverage observations can significantly bias the retrieved chemistry - but equally overfitting can also produce problems. Explicit calculation of information content, such as that featured by \cite{howe17}, may prove useful both for observation planning and also for tailoring retrieval models. 

There are of course several aspects of exoplanet spectral inversion that we have not touched on. Perhaps one of the most significant is the completeness and accuracy of the gas absorption information that is included in retrieval schemes. \cite{tennyson18} provide a summary of the ExoMol project, which is one of the current community efforts to ensure that gas absorption data are as accurate as possible. In addition, the processing of these data for inclusion in retrieval models is also an important step that can be a potential source of error.

Finally, there are other methods for extracting spectral information of exoplanet atmospheres which we have not discussed here, as they are beyond the scope of this paper. These include high-spectral-resolution observations, which can also be used to recover information about exoplanet chemistry and atmospheres (e.g. \citealt{demooij09,schwarz15,hoeijmakers18}); their use in retrieval scenarios is currently being explored \citep{brogi19}. We have also focused on transiting exoplanets in this work; with the launch of \textit{JWST}, and first-light for next generation ground-based telescopes such as the \textit{Extremely Large Telescope} fast approaching, significant advances in direct spectral imaging of exoplanets may also be expected over the current state-of-the-art (represented by e.g. \cite{macintosh15,bonnefoy16,gravity19}), opening up further opportunities to characterise non-transiting worlds.

\begin{acknowledgements}
JKB was supported by a Royal Astronomical Society Research Fellowship while this work was taking place. KH thanks the Swiss National Science Foundation, PlanetS National Center of Competence in Research, European Research Council via Consolider Grant number 71620 and MERAC Foundation for partial financial support. We thank the two anonymous reviewers whose comments improved the clarity of this manuscript.
\end{acknowledgements}
\bibliographystyle{aps-nameyear}      
\bibliography{issi_retrievals.bib}                

\end{document}